\documentstyle[preprint,aps]{revtex}
\begin{document}

\title{Surface melting of methane and methane film on magnesium oxide}
\draft
\tightenlines 
\author{N. Klipa}

\address{Department of Physics, University of Zagreb, Bijeni{\v c}ka 32, 10000 Zagreb, Croatia} 

\author{ G. Bilalbegovi\'c }

\address{Department of Physics, University of Rijeka, Omladinska 14, 51000 Rijeka, Croatia}

\date{to appear in Surface Science}

\maketitle

\begin{abstract}

Experiments on surface melting of several organic materials have
shown contradictory results. We study the Van der Waals interactions between interfaces in
surface melting of the bulk $CH_{4}$ and
interfacial melting of the $CH_{4}$ film on the $MgO$ substrate. 
This analysis is based on the
theory of Dzyaloshinskii, Lifshitz, and Pitaevskii for dispersion forces
in materials characterized by the frequency dependent dielectric functions.
These functions for magnesium oxide and methane 
are obtained from optical data using an oscillator
model of the dielectric response. 
The results show that a repulsive interaction between the solid-liquid and 
liquid-vapor interfaces exists for the bulk methane.
We also found that the van der Waals forces between two solid-liquid interfaces
are attractive for the $CH_4$ film on the $MgO$ substrate.
This implies that the van der Waals forces induce the presence of 
complete surface melting for the bulk methane and
the absence of interfacial melting for $CH_4$ on the MgO substrate.


\end{abstract}

\section{Introduction}

As the temperature is increased from below the bulk triple point, 
some solids begin to melt from their surfaces inward. If the thickness of the
melted layer increases without limit as the temperature approaches the bulk
triple point, then the solid is said to undergo surface melting
\cite{Amsterdam,Dash}. The melting is incomplete if the thickness of the
melted film remains finite as the temperature increases at all temperatures
including the triple point. A special case of incomplete surface melting 
where the liquid film thickness is zero is sometimes referred to as surface 
nonmelting. For example, it was found that the $Pb$(110) surface melts. 
The $Pb$(111) surface does not melt, whereas $Pb$(100) exhibits
incomplete melting. Therefore, incomplete or blocked melting appears
on $(100)$ fcc metal surfaces with an intermediate packing density
between $(110)$ and $(111)$ \cite{Frenken,Goranka}. It is characterized
by a quasi-liquid film of the finite and small thickness.
Surface melting was studied for several metals and other 
materials, such as ice, germanium, and multilayers 
of gases adsorbed on graphite \cite{Amsterdam,Dash}. 
High-temperature disordering for surfaces 
of several organic materials was also studied. 

X-ray reflectivity measurements were used for high-temperature studies of
the $(100)$ face of caprolactam ($C_6H_{11}ON$) \cite{Chandavarkar,Wong}.  
Initially a series of layering transition followed by 
prewetting was suggested, but later 
a surface hydration mechanism was proposed for temperature disordering of
this surface. Chernov and Yakovlev
studied surface melting for the $(001)$ and $(010)$ faces of biphenyl 
($C_{12}H_{10}$) using the ellipsometry technique \cite{Chernov}. 
They found that the thickness of a quasi-liquid layer at $0.5$ K
below the bulk melting point reaches $100 \AA$ for the $(001)$ face 
and $300 \AA $ for the $(010)$ face. In contrast, de Jeu and coworkers 
using the synchrotron X-ray reflectivity measurements have shown the
absence of surface melting for the $(001)$ face of the bulk biphenyl 
up to $2$ K below the melting point \cite{deJeu}. It was proposed that surface
melting observed 
by Chernov and Yakovlev is induced by the
roughness of the glass used in their ellipsometry experiment \cite{deJeu}.
Surface properties of biphenyl, caprolactam, and other
organic materials and organic thin films at high temperatures
deserve further experimental and theoretical studies.
Investigations of organic films are also important for applications,
such as tribology and lubrication, corrosion inhibition, modification of 
the substrate optical properties, molecular electronics, and various 
industrial chemical processes.

Surface melting for a $\sim 10$ layer thick film of $CH_4$ adsorbed on
$MgO(100)$ was studied by neutron scattering
\cite{Bienfait,Gay}. It was found that the thickness of the quasi-liquid phase 
changes from one layer at $72$ K to $\sim 6$ layers at $90.3$ K. 
(At atmospheric pressure
the bulk melting point of methane is $90.66$ K, whereas the liquid-vapour 
transition occurs at $111.56$ K.) The neutron-scattering measurements of the
translational diffusion coefficient above $72$ K
have shown that this quantity is in the
$10^{-5}$ $cm^2$ $s^{-1}$ range, i.e., it has a value typical for mobility
in quasi-liquid films. 
It was found that for a methane thick
film ($\sim 20$ layers) at $T=94$ K (i.e., above the bulk melting temperature)
almost all layers are liquefied \cite{Gay}. 
Recently the layering and melting properties of methane on $MgO(100)$
were studied between $70$ K and $96$ K using high-resolution 
adsorption isotherms and for the film thickness of up to five layers.
\cite{Larese}. Melting transition was observed at $80$ K for the monolayer,
and at $85$ K for layers $2$ through $4$.

Related to the  experiments on surface melting of bulk 
caprolactam and biphenyl, it would be interesting to study surface melting of
a bulk methane. These studies should determine whether surface melting of
the $CH_4$ film on $MgO$ is inherent feature of this organic material,
or it is induced by the substrate.
Interfacial melting is an appearance of a liquid film at the interface between
one solid material and the substrate of the other material as the bulk melting point
of the first material is approached \cite{Wilen,Beaglehole}.
Therefore, surface melting is a special case of 
interfacial melting where the substrate is the vapor phase of the first material. 
In this work long-range interactions important for surface melting of a bulk methane, 
as well as  for interfacial melting of a methane film on the MgO substrate
are analyzed using the theory of Dzyaloshinskii, Lifshitz, and Pitaevskii
(DLP) \cite{Dlp}. Our findings revealed that complete surface melting 
exists for the bulk methane.
Because of an attractive interaction between interfaces 
melting is absent at the boundary between 
the $CH_4$ film and the $MgO$ substrate.

In the following  
the DLP theory of the van der Waals interaction between planar
interfaces and its consequences for surface melting are described in Sec. II. 
The calculation of frequency dependent dielectric functions  of $CH_4$ and $MgO$ is 
discussed in Sec. III.
Results and discussion for the surface and interfacial melting problems 
are presented in Sec. IV. 
A summary and conclusions are given in  Sec. V.

\section{Surface melting and Van der Waals forces}

The surface free energy per unit area for a system with a quasi-liquid film
of the thickness $l$, at the temperature $T$ is 
\cite{Amsterdam,Dash,Pluis}
\begin{equation}
{\cal F}(l,T)=\gamma_{sl} + \gamma_{lv} + L_m (1-T/T_m)l +
F_{sr} + F_{lr}.
\label{eq:1}
\end{equation}
In this equation $\gamma_{sl}$ and $\gamma_{lv}$ are the free energies of
the solid-liquid and liquid-vapor interfaces, respectively, $L_m$ is the
latent heat of melting per unit volume, and $T_m$ is the bulk melting
temperature. The fourth term $F_{sr}$ represents the gain in interfacial
energy associated with the wetting of the solid surface by a quasi-liquid
film. This term is the result of short-range interactions. 
Their contribution is oscillatory and decreases exponentially to zero
for large $l$. Short-range forces produce a logarithmic temperature dependence 
of the liquid film thickness as the bulk melting temperature is approached.
Long range van der Waals interactions are given by the
last term $F_{lr}$ in Eq. (1). 
The contribution of the van der Waals forces for 
non-retarded interactions (where $l$ is much smaller than the wavelength
of light for a characteristic excitations) is given by $H l^{-2}$.
The parameter $H$ is the Hamaker constant and gives information about the 
magnitude of the van der Waals interactions in different media
\cite{Israelachivili}. (Sometimes $H'=12\pi H$  or $H''=-12\pi H$ are
called
the Hamaker constant.)  
Surface melting occurs when $H>0$, i.e., when two interfaces 
(solid-liquid and liquid-vapour) repel each other.
Then  $F_{lr}(l)$ is monotonously decreasing function, or its minimum is
not deep enough to bind the interface.
The van der Waals contribution produces a power-law temperature dependence of the
liquid film thickness as $T\rightarrow T_m$ 
\begin{equation}
l=\left[\frac {L_m(T_m-T)}{2HT_m}\right]^{-1/3}.
\label{eq:2}
\end{equation} 
For $H<0$ two interfaces attract each other and the melting is at most
incomplete. Then a deep minimum exists on the free energy 
curve $F_{lr}$, and the interface is trapped there.

Dzyaloshinskii, Lifshitz, and Pitaevskii  developed the theory of
long-range dispersion interaction between macroscopic bodies 
\cite{Dlp,Israelachivili}.  
The van der Waals interaction between interfaces 
is a consequence of electromagnetic fluctuations in a polarizable medium
and the discontinuity of the dielectric function across the boundaries.
In the DLP theory the force between interfaces 
(such as those shown in Fig. 1) is a function of the frequency-dependent 
dielectric functions $\epsilon(\omega)$ of materials in contact.
In the following all media are taken as isotropic and continuous. 
The free energy per unit area of two media (1 and 2) separated by
a film of thickness $l$ (media 3, see Fig. 1) is \cite{Dlp,Elbaum,Wilen}
\begin{eqnarray}
F_{lr}(l,T)= \frac {kT}{8\pi l^2}
\sum_{m=0}^{\infty} \ '
\int_{r_{m}}^{\infty} 
dx \ x \ (ln (1- \frac {(x-x_1)(x-x_2)} {(x+x_1)(x+x_2)} 
exp(-x)) \nonumber \\
+ln (1- \frac {(\epsilon _2 x -\epsilon _3 x_2)
(\epsilon _1 x - \epsilon _3 x_1)}
{(\epsilon _2 x + \epsilon _3 x_2)(\epsilon _1 x + \epsilon_3 x_1)} exp(-x))),
\end{eqnarray}
where 
\begin{equation}
r_m=2l\xi _m (\epsilon _ 3)^{1/2}/c,
\label{eq:4}
\end{equation}
and 
\begin{equation}
x_{1,2}= \left(x^2 - r_m^2(1- \frac {\epsilon_{1,2}} 
{\epsilon_3})\right)^{1/2}.
\label{eq:5}
\end{equation}
The dielectric functions $\epsilon _1 (\omega)$,
$\epsilon _2 (\omega)$, and $\epsilon _3 (\omega)$ are evaluated at
imaginary frequencies $i\xi _m = i(2\pi k T/{\hbar})m$. 
The functions $\epsilon(i\xi)$ are positive, real and
decrease monotonically from the static dielectric function $\epsilon _0$
for $\xi=0$, to $1$ as $\xi \to \infty$. 
In the above expressions $k$, $\hbar$, and $c$ are 
the Boltzmann and Planck constants and velocity of light, respectively.
The prime on the sum in Eq. (3)
means that the term $m=0$ has to be multiplied with 
$\frac {1} {2}$.
In the derivation of Eq. (3)  retarded interactions were assumed. 
If retardation is ignored,
then $c \to \infty$, $r_m \to 0$, and Eq. (1) simplifies to
\begin{equation}
F_{lr}^{nr}(l,T)= \frac {kT} {8\pi l^2} \sum_{m=0}^{\infty}
\int_0^{\infty} 
dx \ x \ ln \left(1- \frac {(\epsilon _2 (i\xi_m) -\epsilon _3(i\xi_m) )
(\epsilon _1 (i\xi_m) - \epsilon _3(i\xi_m) )}
{(\epsilon _2 (i\xi_m) + \epsilon _3(i\xi_m) )
(\epsilon _1(i\xi_m)  + \epsilon_3(i\xi_m))} exp(-x) \right).
\label{eq:6}
\end{equation}
A common approach 
is to obtain an approximate value of $\epsilon (\omega)$
from the optical measurements
\cite{Wilen,Israelachivili,Elbaum,Chen,DalCorso,Parsegian,Hough}. 
An oscillator model of the dielectric response is then used to  
generate data, i.e.,
the dielectric function is represented by 
\begin{equation}
\epsilon (\omega) = 1 + \sum_i \frac {f_i} {{e_i}^2 - i\hbar\omega g_i
-(\hbar\omega)^2},
\label{eq:7}
\end{equation}
where $f_i$, $e_i$, and $g _i$ are fitting parameters.
The DLP theory was recently applied to surface
melting of ice \cite{Elbaum,Wilen}, germanium, and several metals 
\cite{Chen,DalCorso}.
Unfortunately, optical and dielectric  properties of many organic materials
are less studied than, for example of metals, semiconductors, or
$H_2O$.
As a consequence, we were not able to find necessary optical data for 
biphenyl and caprolactam, two organic materials for which 
experiments on surface melting were performed. In contrast, 
we found data for optical properties
of organic material $CH_4$ and dielectric substrate $MgO$ \cite{Martonchik,Palik}.

\section{Dielectric functions of methane and MgO}

The complex dielectric functions $\epsilon(i\xi)$ were constructed
using experimental data for optical constants of $CH_4$ and $MgO$ 
\cite{Martonchik,Palik}.
As suggested in Ref. \cite{Elazar} we minimized the objective function
\begin{equation}
E= \sum_{j=1}^N 
{\left[\left |{\frac
{\epsilon_r(\omega_j)-\epsilon_r^{exp}(\omega_j)}
{\epsilon_r^{exp}(\omega_j)}}\right | +
\left | {\frac 
{\epsilon_i (\omega_j) - \epsilon_i^{exp} (\omega_j)}
{\epsilon_i^{exp}(\omega_j)}}\right |\right]}^2,
\label{eq:8}
\end{equation}
where $\epsilon_r(\omega)$ and $\epsilon_i(\omega)$ are the real and 
imaginary parts of a dielectric function given by Eq. (7), and '{\it exp}'
denotes experimental values taken from Refs. \cite{Martonchik,Palik}.
Therefore, the calculation was performed by finding a minimum of
an objective function based on  
a simultaneous fitting of both real and imaginary parts of the
dielectric function. We found that, in comparison with some other functions,
the functional form in Eq. (8) is  suitable for minimization.
In principle, the most important factor in the minimization process is the
quality of experimental data, i.e., the number of points and their 
distribution. It is known that the main contribution belongs to  data in the UV
region of the spectra \cite{Parsegian,Hough}.
The UV part of the spectrum for $MgO$ is represented with many points
\cite{Palik}. In contrast, in the UV spectra for the solid and liquid methane 
few points exist \cite{Martonchik}.
A smaller number of points gives more freedom in the choice of the parameters
which minimize Eq. (8). Therefore, more local minima exist and they
produce similar values of the objective function. 
This is a limitation for the calculation of dielectric properties of methane.
Fortunately previous calculations for other
materials have shown that the complex dielectric functions 
$\epsilon(i\xi)$ which enter expressions in the DLP theory are not
very sensitive to the fitting procedure and
the detailed experimental information. 
As suggested in Ref. \cite{Parsegian} and followed by others
\cite{Elbaum,Wilen}, we used the constraint $\epsilon_r=n_{vis}^2$,
where $n_{vis}$ is the refractive index in the visible part of the
spectrum. The value $n_{vis}$ is known with a great accuracy and
fitting with this constraint gives better results. 
Objective functions were minimized using the simplex method
\cite{Numrec,Multisimplex}.
The accuracy of calculation was checked by changing minimization routines and
also by calculating dielectric functions for other materials
where extensive data are available 
\cite{Wilen,Israelachivili,Elbaum,Chen,DalCorso,Parsegian,Hough}.

Martonchik and Orton,
motivated by astrophysical applications, recently
calculated the optical constants of liquid and solid methane using the
spectra from the literature \cite{Martonchik}. Liquid methane was
studied at the boiling and melting temperatures, whereas the solid phase I
of methane was studied at the melting point and at $T=30$ K.
(The solid phase I is the face centered cubic phase of methane which 
at atmospheric pressure exists for $T>20$ K.)
Therefore, data obtained by Martonchik and Orton for the solid and 
liquid methane at the melting point were used to model the solid and liquid 
phase. For the vapor phase we put $\epsilon=1$. 
The static dielectric function of the gas methane is $\epsilon=1.000944$ \cite{CRC}.
The fits for $CH_4$ were performed
to match dielectric functions at zero frequency. These 
static dielectric functions  
are $\epsilon_0=1.67$ for the liquid methane, and
$\epsilon_0=1.74$ for the solid methane \cite{Martonchik}.
The calculated complex dielectric functions of the solid and liquid methane 
are shown in Fig. 2. 
Corresponding fitting parameters are given in Tables I and II. 
The curves $\epsilon (i\xi)$ for the solid and liquid methane are close to
each other, but do not cross. In similar calculation for surface melting
of ice it was found that the complex dielectric functions of
water and ice 
cross at $10$ eV (i.e., for $m \sim 205$) \cite{Elbaum,Wilen}.

The detailed review of optical properties of $MgO$ was
given by Roessler and Huffman \cite{Palik}.
The values of optical constants were tabulated in the range $(0.002-586)$ eV.
Because of a large set of data we use only selected points.
From these data 
the function $\epsilon(i\xi)$ for $MgO$ was constructed. 
The static dielectric function of $MgO$ is 
$\epsilon_0 = 9.8$ \cite{Palik}. 
Here, as in Ref. \cite{Sabisky} where the fitting 
based on a double band was performed, 
we did not match the static dielectric function.
The value of static dielectric function of $MgO$ is much larger than the
average value of $\epsilon_r$ in the UV part of the spectrum.
Additional problem is the existence of a close sharp peak at
$\hbar \omega = 7.8 $ eV. Therefore,
the inclusion of the static dielectric function into the fit for $MgO$ 
prevents a good behavior of the whole fitted curve.
Fitting parameters for $MgO$ are given in Table III. 
The complex dielectric functions $\epsilon(i\xi)$ of 
$MgO$  is shown in Fig. 2.
Our four-band result is almost the same as a double band calculation 
from Ref. \cite{Sabisky}.

\section{Surface and interfacial melting}

\subsection{Bulk methane}

In surface melting problem  
the frequency dependent dielectric functions of solid, 
liquid (melt), and gas methane are respectively:
$\epsilon _1 = \epsilon _s (\omega)$,
$\epsilon _3 = \epsilon _l (\omega)$, and 
$\epsilon _2 = 1$ (see Fig. 1).
The integral in Eq. (3) was calculated numerically using the Gaussian method.
The accuracy of the integration technique was checked by changing 
numerical procedures and also by reproducing the results for ice 
\cite{Elbaum}.  
The calculated free energy as a function of liquid film thickness is 
shown in Fig. 3. The minimum is reached for $l\rightarrow\infty$. 
This functional form of $F(l)$ is typical for complete surface melting. 
We found that the Hamaker constant is 
$H=5.26 \times 10^{-22}$ J. Similar values were calculated for metals with
melted surfaces, such as $Pb$, $Al$, and $Au$ \cite{Chen}.

\subsection{Methane film on MgO}

In interfacial melting problem,
and for interfaces shown in Fig. 1: 
$\epsilon_1 = \epsilon_{MgO}(\omega)$
is dielectric function for $MgO$,
$\epsilon_2 = \epsilon_s (\omega)$ is for a solid phase, 
and $\epsilon_3 = \epsilon_l (\omega)$ is for a liquid phase of methane.
Therefore, we consider the solid $CH_4$ - liquid $CH_4$ - $MgO$ substrate
system. The same numerical procedure as for the surface melting problem was applied.
The free energy is shown in Fig. 4. The Hamaker constant is 
$H=-6.03 \times 10^{-22}$ J.
Therefore, the van der Waals forces produce an attractive interaction 
between interfaces and prevent interfacial melting. 
The functional form of $F(l)$ shown in Fig. 4 was not found for interfacial
melting of ice on various substrates \cite{Wilen}, but two other forms
of $F(l)$ were calculated there. They correspond to complete and
incomplete interfacial melting of ice on the particular substrate.

\section{Summary and Conclusions}

The interaction of interfaces present in  
surface melting of the bulk methane and interfacial melting of 
the $CH_4$ film on the $MgO$ substrate
is analyzed within the
Dzyaloshinskii, Lifshitz, and Pitaevskii theory of 
the van der Waals interaction. Frequency dependent dielectric functions for
these materials are calculated from optical data.
It was found that complete surface melting exists for methane.
Our study is the first investigation of surface melting properties of the
bulk methane. 
We also found that the van der Waals interactions do not produce 
interfacial melting of the $CH_4$ film on the $MgO$ substrate.
Therefore, premelting observed in the experimental investigations
of the $CH_4$ films on the $MgO$ substrate \cite{Bienfait,Gay,Larese}
is inherent feature of the bulk methane.
Methane is till now the only  organic material for which clear
complete surface melting was found.
The results obtained here depend on the quality of the
optical measurements used to model dielectric functions.
Nevertheless, these results shed light on the
interaction between interfaces in surface and interfacial melting involving
methane. These results suggest 
experimental studies of surface melting of the bulk methane, 
for example using the synchrotron X-ray reflectivity.
In addition, {\it ab initio} molecular dynamics study of 
the high-temperature properties 
of methane surfaces is feasible, 
similar as a recent simulation of a high-pressure behavior for
bulk $CH_4$ \cite{Francesco}.
It is important to do all necessary optical measurements for biphenyl
and caprolactam, and then the calculation of the van der Waals interactions for
these materials.
There is a possibility that
surface melting observed for biphenyl on the glass surface \cite{Chernov} 
is interfacial melting induced by the dielectric properties of the substrate. 
Till now the roughness of the glass and impurity induced 
interfacial melting were
proposed to explain the results of Chernov and Yakovlev. 
A possibility of interfacial melting induced by the van der Waals forces
for biphenyl on the glass 
also explains the absence of surface melting 
for the bulk biphenyl found using the X-ray reflectivity \cite{deJeu}.
Recent atomic force microscopy study of the (001) surfaces of 
$n-C_{23}H_{48}$ paraffin crystals \cite{Plomp} shows that this technique
gives a good opportunity to investigate melt growth on a molecular scale
for surfaces of methane and other organic materials.

\acknowledgments
This work has been carried out under the HR-MZT project 119206 -
``Dynamical Properties of Surfaces''.
We would like to thank I. Kup{\v c}i{\' c} and E. Ljubovi{\'c}
for their help.

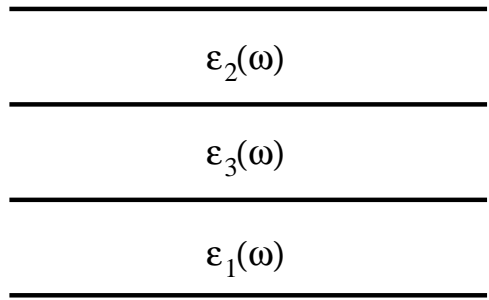
\begin{figure}
\caption{Schematic representation for configuration of interfaces between
media characterized by the frequency
dependent dielectric functions.}
\label{fig1}
\end{figure}

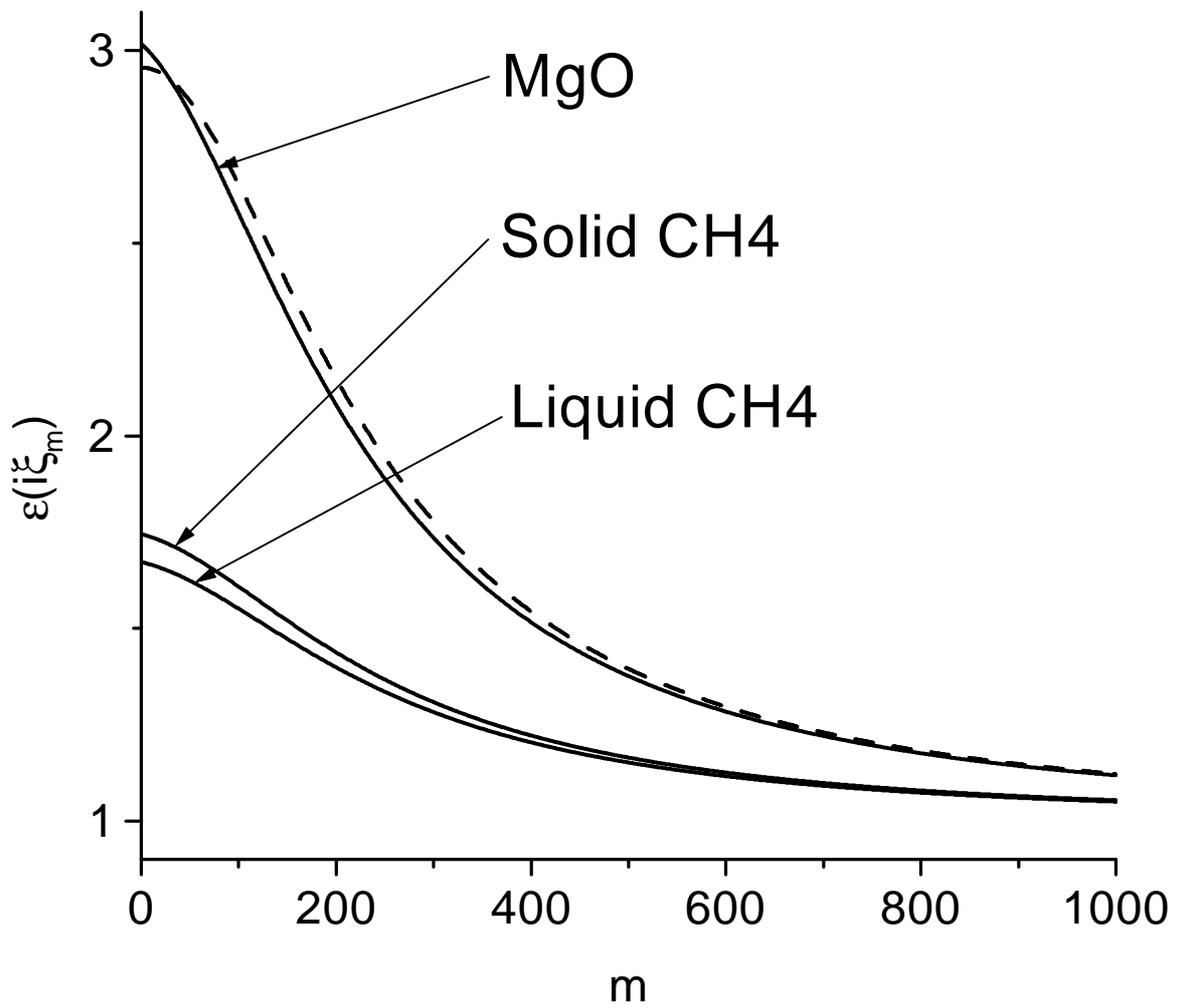
\begin{figure}
\caption{Complex dielectric functions. Dashed line  for $MgO$ is the fit
from Ref. [28]. The index $m$ is defined in Eqs. (3) and (6)
in the text, and $m=1000$ corresponds to $48.8$ eV. The contribution of
various $m$-dependent terms in Eqs. (3) and (6) is obvious.}
\label{fig2}
\end{figure}

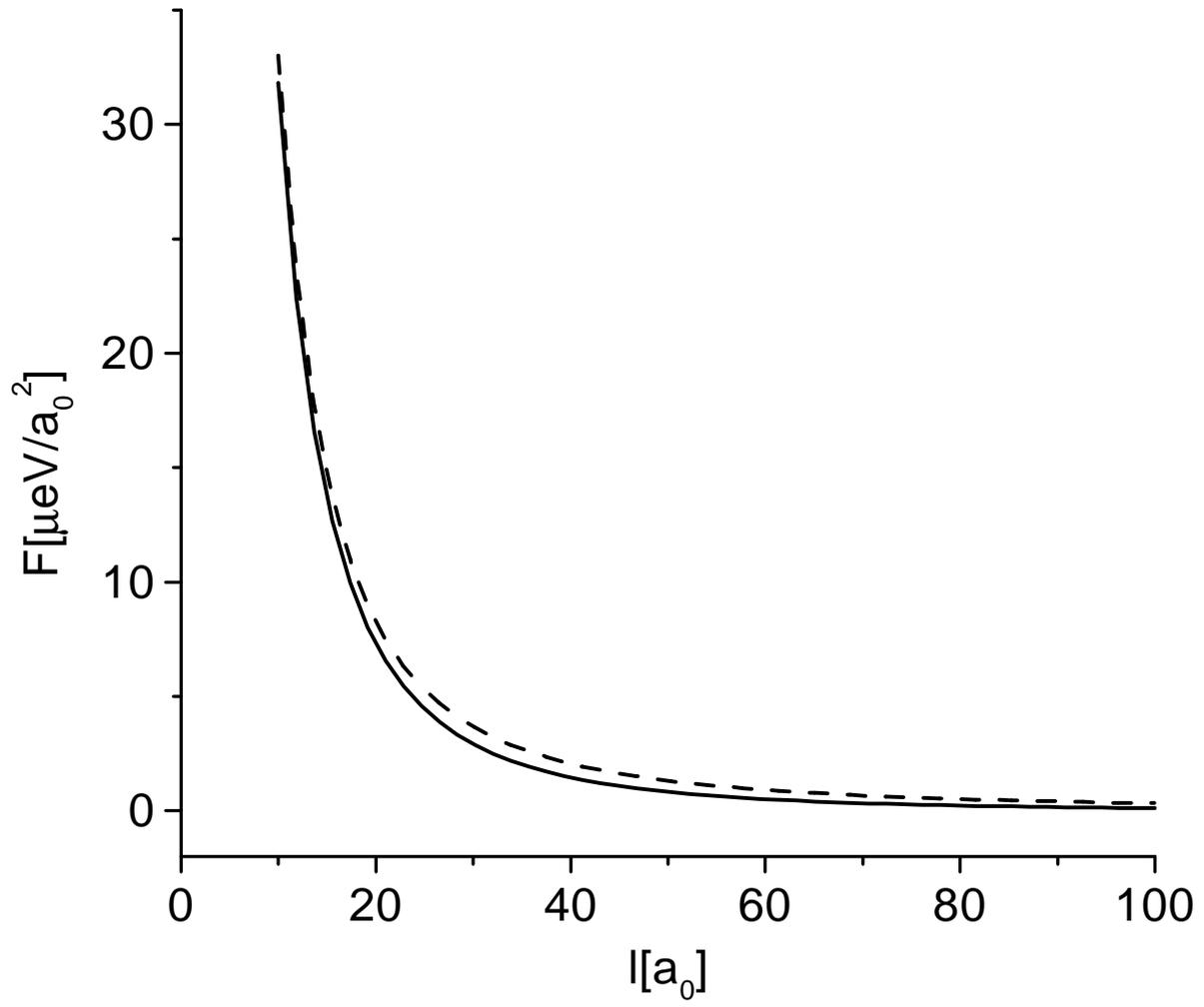
\begin{figure}
\caption{The free energy per unit area ($a_0$ is the Bohr radius)
as a function of the film thickness $l$
for surface melting of the bulk methane.
The dashed line shows the free energy without retardation,
whereas the full line represents the result with retardation.}
\label{fig3}
\end{figure}

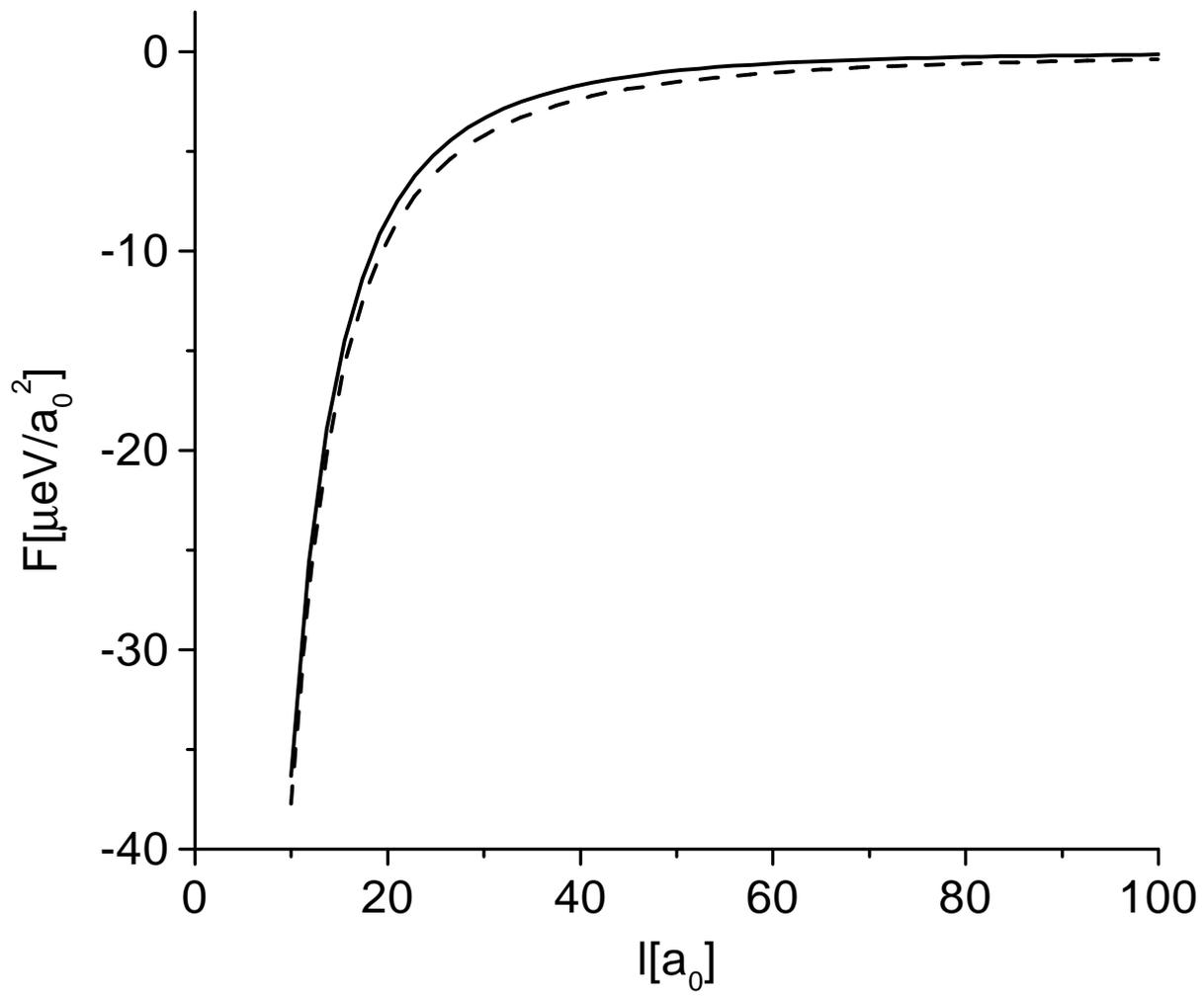
\begin{figure}
\caption{The free energy vs film thickness 
for interfacial melting of a methane film on $MgO$. 
Details as in Fig. 3.}
\label{fig4}
\end{figure}

\begin{table}
\caption{Parameters used for an oscillator model of dielectric response
for the solid methane using data from Ref. [22].}
\label{table1} 
\begin{tabular}{l l l } 
$e_i$ (eV)  & $f_i$ ($eV^2$) & $g_i$ (eV)\\ 
\hline
$10.0$  & $10.0$ & $1.8$ \\ 

$12.7$  & $44.0$ & $3.6$ \\ 

$14.4$  & $64.0$ & $3.5$ \\ 

$22.3$   & $10.0$ & $2.4$ \\

$23.1$  & $23.0$ & $6.8$ \\ 
\end{tabular}
\end{table}

\begin{table}
\caption{Parameters used for an oscillator model of dielectric response
for the liquid methane using data from Ref. [22].}
\label{table2} 
\begin{tabular}{l l l } 
$e_i$ (eV)  & $f_i$ ($eV^2$) & $g_i$ (eV) \\ 
\hline
$10.0$  & $8.0$ & $1.5$ \\ 

$12.8$  & $42.0$ & $3.5$ \\ 

$14.8$  & $62.0$ & $4.0$ \\ 

$22.7$   & $8.0$ & $2.4$ \\

$23.8$  & $21.0$ & $6.5$ \\ 
\end{tabular}
\end{table}

\begin{table}
\caption{Parameters used for an oscillator model of dielectric response
for solid MgO using data from Ref. [23].}
\label{table3} 
\begin{tabular}{l l l } 
$e_i$ (eV)  & $f_i$ ($eV^2$) & $g_i$ (eV) \\ 
\hline
$7.6$  & $3.0$ & $0.1$ \\ 

$10.8$  & $150.0$ & $3.5$ \\ 

$13.2$  & $40.0$ & $1.0$ \\ 

$17.0$   & $130.0$ & $4.0$  \\
\end{tabular}
\end{table}

\end{document}